\documentclass[dvips]{arxstspdf}
\usepackage{flushend}
\usepackage{stfloats}
\usepackage{graphicx}
\volume{26}
\issue{2}
\pubyear{2011}
\firstpage{212}
\lastpage{226}
\doi{10.1214/10-STS331}

\makeatletter
\newproclaim{example}{Example}
\makeatother

\begin{document}
\begin{frontmatter}

\title{Bayesian Models and Methods in Public Policy and Government
Settings\thanksref{T1}}
\relateddoi{T1}{Discussed in \doi{10.1214/11-STS331A},
\doi{10.1214/11-STS331B}
 and \doi{10.1214/11-STS331C}; rejoinder at \doi{10.1214/11-STS331REJ}.}
\runtitle{Bayesian Methods in Public Policy}
\begin{aug}
\author[a]{\fnms{Stephen E.} \snm{Fienberg}\corref{}\ead[label=e1]{fienberg@stat.cmu.edu}
\ead[label=u1,url]{http://www.stat.cmu.edu/fienberg/}}
\runauthor{S. E. Fienberg}

\affiliation{Carnegie Mellon University}

\address[a]{Stephen E. Fienberg is Maurice Falk University Professor,
Department of Statistics, Heinz College, Machine Learning Department, and Cylab,
 Carnegie Mellon University, Pittsburgh, Pennsylvania 15213-3890, USA
\printead{e1,u1}.}

\end{aug}

% ABSTRACT
%
\begin{abstract}
Starting with the neo-Bayesian revival of the 1950s, many statisticians
argued that it was inappropriate to use Bayesian methods, and in
particular subjective Bayesian methods in governmental and public
policy settings because of their reliance upon prior distributions. But
the Bayesian framework often provides the primary way to respond to
questions raised in these settings and the numbers and diversity of
Bayesian applications have grown dramatically in recent years. Through
a series of examples, both historical and recent, we argue that
Bayesian approaches with formal and informal assessments of priors AND
likelihood functions are well accepted and should become the norm in
public settings. Our examples include census-taking and small area
estimation, US election night forecasting, studies reported to the
US Food and Drug Administration, assessing global climate change, and
measuring potential declines in disability among the elderly.
\end{abstract}

% KEYWORDS
%

\end{frontmatter}

%s1 ###
\section{Introduction and History}\label{sec1}

Beginning with the posthumous publication in \citeyear{Bayes1763} of the essay
attributed to the Rev. Thomas Bayes, and continuing well into the
twentieth century, virtually the only approach to statistical inference
was the method of inverse probability based on applications of Bayes's
theorem (see, e.g., Fienberg, \citeyear{Fienberg2006a}). Nonetheless,
most applications of
statistical methods in governmental settings were based primarily on
descriptive statistics and there was little debate regarding the
relevance of Bayesian approaches in public life despite efforts at
implementation, for example, Laplace's development of ratio estimation
to estimate the size of the population of France.

Criticism of the method of inverse probability, as Bayesian methodology
was known for almost\break  200~years, began in the mid-19th century with the
rise of a philosophical school advocating objective probability. The
fundamental concern of the objectivists was the requirement for a prior
distribution and they argued for a frequentist view of probability.
Unfortunately they failed to present a methodology for inference to
counter that of inverse probability and it was not until the work of R.
A. Fisher and Jerzy Neyman and Egon Pearson in the 1920s that serious
alternative statistical procedures were in place. Neyman's (\citeyear{Neyman1934})
critique of Gini's version of the representative method for survey
taking not only ushered the frequentist repeated sampling perspective
into the realm of official statistics, but it also introduced the
frequentist tool of confidence intervals and its long-run repeated
sampling interpretation (see Fienberg and Tanur, \citeyear{Fienberg1996}).

%It is especially appropriate for me to be presenting this talk in
%Italy because of the importance of the work of Bruno de Finetti who in
%the 1930s developed the theoretical arguments in support of subjective
%probability. See especially de Finetti (\citeyear{Finetti1937}).
Bayesian tools played an important role in a number of statistical
efforts during World War II, including Alan Turing's work at Bletchley
Park, England, to crack the Enigma code, but with the creation of such
frequentist methods as sequential analysis by Barnard in England and
Wald in the United States and the elaboration of design-based analyses
in sample surveys, as statistics passed the mid-century mark,
frequentist approaches were in the ascendancy in the public arena. This
was especially true in statistical agencies where the ideas of random
selection of samples and repeated sampling as the basis of inference
were synonymous, and statistical models and likelihood-based methods
frowned upon at best.

With the introduction of computers for statistical calculations in the
1960s, however, Bayesian methods began a slow but prolonged comeback
that accelerated substantially with the introduction of
Mar\-kov chain Monte Carlo (MCMC) methods in the early 1990s. Today Bayesian methods
are challenging the supremacy of the frequentist approaches in a wide
array of areas of application.

How do the approaches differ? In frequentist infe\-rence, tests of
significance are performed by suppo\-sing that a hypothesis is true (the
null \mbox{hypothesis}) and then computing the probability of observing
a~sta\-tistic at least as extreme as the one actually~%
ob\-served during hypothetical future repeated trials conditional on the
parameters, that is, a $p$-va\-lue. Baye\-sian inference relies upon direct
inferences about parameters or predictions conditional on the
observations. In other words, frequentist statistics examines the
probability of the data given a~mo\-del (hypothesis) and looks at
repeated sampling properties of a procedure, whereas Bayesian
statistics examines the probability of a model given the observed data.
Bayesian methodology relies largely upon Bayes's theorem for computing
posterior probabilities and provides an internally consistent and
coherent normative methodology; frequentist methodology has no such
consistent normative framework. Freedman (\citeyear{Freedman1995}) gave an overview of
these philosophical positions, but largely from a frequentist
perspective that is critical of the Bayesian normative approach.

The remainder of the article has the following structure. In the next
section I give a summary of some of the most common and cogent
criticisms of the Bayesian method, especially with regard to its use in
a public context. Then in Section~\ref{sec3}, through a series of examples, both
historical and recent, I~argue that Bayesian approaches with formal and
informal assessments of priors and likelihood functions are
well accepted and should become the norm in public settings. My
examples include US election night forecasting, census-taking and
small area estimation, studies reported to the US Food and Drug
Administration, assessing global climate change, and measuring declines
in disability among the elderly. We conclude with a brief summary of
challenges facing broader implementation of Bayesian methods in public contexts.

I do not claim to be providing a comprehensive account of Bayesian
applications but have merely attempted to illustrate their breadth. One
area where Bayesian ideas have made serious inroads, both in theory and
in actual practice, but which we do not discuss here is the law (e.g.,
see Fienberg and Kada\-ne, \citeyear{Fienberg1983}; Donnelly, \citeyear
{Donnelly2005}; Taroni et al., \citeyear{Taroni2006};
Kadane,  \citeyear{Kadane2008}). The present article includes a purposeful selection of
references to guide the reader to some of the relevant recent Bayesian
literature on applications in the domains mentioned, but the list is
far from comprehensive and tends to emphasize work closest to my own.

%s2 ###
\section{The Arguments For and Against the Use of Bayesian Methods}\label{sec2}

\textit{Bayesian and frequentist inference in a nutshell:} It is
especially convenient for the present purposes to think about Bayes's
theorem in terms of density functions. Let $h(y | \theta)$ denote the
conditional density of the random variable $Y$ given a parameter value~%
$\theta$ in the parameter space~$\Theta$. Then we can go from the prior
distribution for $\theta$, $g(\theta)$, to that associated with~$\theta
$ given $Y=y$, $g(\theta| y)$, by
%
%
%e2 ###
%e1 ###
\begin{eqnarray}\label{eq1}
g(\theta| y) &=& h(y | \theta) g(\theta)\big / \sum_{\theta\in\Theta} h(y
| \theta) g(\theta)\\
\eqntext{\mbox{if } \theta \mbox{ has a discrete
distribution,}}\\\label{eq2}
g(\theta| y)&=& h(y | \theta) g(\theta) \big / \int_{\Theta} h(y | \theta)
g(\theta)\,d\theta\\
\eqntext{\mbox{ if } \theta\mbox{ has a continuous distribution.}}
\end{eqnarray}

$\!\!$Bayesians make inferences about the parameters~by looking directly at
the posterior distribution $g(\theta|y)$ given the data $y$.
Frequentists make inferences\break about $\theta$ indirectly by considering
the repeated sampling properties of the distribution of the data $y$
given the parameter $\theta$, that is, through $h(y|\theta)$. Baye\-sians
integrate out quantities not of direct substantive interest and then
are able to make probabilistic inferences from marginal distributions.
Most frequentists use some form of conditioning argument for inference
purposes while others maximize likelihood functions. Frequentists
distinguish between random variables and parameters which they take to
be fixed and this leads to linear mixed models where some of the effects
are fixed, that is, are parameters, and some are random variables. For
a~Baye\-sian all linear models are in essence random effects models since
parameters  {are} themselves considered as random variables. Thus it
is natural for a Bayesian to consider them to be independent draws from
a~common distribution, $g(\theta)$, that is, treating them as~%
ex\-changeable following the original argument of de Fi\-netti (\citeyear{de1937}). This
approach leads naturally to put\-ting distributions on the parameters of
prior distributions and to what we now call the hierarchical Bayesian
model. It is the normalizing constants [the denominators of (\ref{eq1}) and
(\ref{eq2})] that are notoriously difficult to compute and this fact has led,
in large part, to the use of MCMC methods such as Gibbs sampling that
involve sampling from the posterior distribution.

A reviewer of an earlier version of this article suggested that
hierarchical models are really not Baye\-sian, unless one puts a prior at
the top level of the hierarchy. This ignores history. As Good (\citeyear{Good1965})
noted, his own use of such ideas draws on work dating back at least to
the 1920s and the work of W.~E.~Johnson whose ``sufficientness''
postulate implicitly used finite exchangeable sequences. And while
non-Baye\-sians came to recognize the power of such structures many
decades later they did attempt to emulate the Bayesian approach, but of
course without the clean Bayesian probabilistic interpretation.

\textit{Critique of the Bayesian perspective:} The most common
criticism of Bayesian methods is that, since there is no single correct
prior distribution, $g(\theta)$, all conclusions drawn from the
posterior distribution are suspect. One counter to this argument is
that published analyses using Bayesian methods should consider and
report the results associated with a variety of prior distributions,
thus allowing the reader to see the effects of different prior beliefs
on the posterior distribution of a parameter. Others argue that one
should choose as a prior distribution one that in some sense eliminates
personal subjectivity. Examples of such ``objective'' priors are those
that are uniform or diffuse across all possible values\vadjust{\eject} of the
parameter, or those that are ``informationless.'' Berger (\citeyear{Berger2006}) and
Goldstein (\citeyear{Goldstein2006}) presented arguments in favor of the objective and
subjective Bayesian approaches in a forum followed by extensive
discussion. For a discussion of the fruitlessness of the search for an
objective and informationless prior, see the article by Fienberg (\citeyear{Fienberg2006b}).

There are a number of other features associated with the subjective
approach including the elicitation of information for the formulation
of prior distributions and the use of exchangeability in the
development of successive layers of hierarchical models. A number of
the examples described in the sections that follow utilize subjective
Bayesian features although not always with full elicitation.

One characteristic of Bayesian inference that weakens this criticism of
the reliance on the prior distribution is that
the more data we collect, the less influence the prior distribution
has on the posterior distribution relative to that of the data. There
are situations, however, where even an infinite amount of data may not
bring two people into agreement (see, e.g., Diaconis  and Freedman,  \citeyear{Diaconis1986}).

Another aspect of the Bayesian methodology that arises in many
applications is the manner in which it ``borrows strength'' when we are
estimating many parameters simultaneously, especially through the use
of hierarchical models. This feature, which is usually viewed as a
virtue, has also been the focal point of criticism by frequentists. For
example, see the commentary by Freedman and Navidi (\citeyear{Freedman1986}) in the
context of census adjustment, in which they critiqued a Bayesian
methodology at least in part because it resulted in the use of data
from one state to adjust the census-based population figures in other
ones. Today, borrowing strength via cross-area regression models is
common in frequentist circles, and the Freedman--Navidi argument thus
takes on a nonstatistical legal issue rather than a statistical one.

For an interesting dialog on different frequentist perspectives
related to statistical inference, see the discussion paper by a group
of frequentist statisticians at Groningen University in The
Netherlands, Kardaun et al. (\citeyear{Kardaun2003}), which was a response to a~series of
questions posed by David Cox following a~lecture at Groningen. As
someone else has noted, it is a~rare occasion where frequentists
seriously entertain ideas such as those extolled by de Finetti (\citeyear{de1937})
and attempt to reject them. A number of the questions discussed in this
article arise in the context of the examples that follow.

%s3 ###
\section{Small Area Estimation and Census Adjustment}\label{sec3}

\textit{Small area estimation:} As we have already intimated, small
area estimation has been a ripe area for Bayesian methods although
because so much of the literature has been oriented toward national
statistical agency problems, the area is dominated by frequentist
techniques and assessments. Surveys conducted by national statistical
agencies typically generate ``reliable'' information either at national
or regional levels. But the demand for information at lower levels of
disaggregation is sufficiently great and resources tend to be
relatively scarce, so that techniques that bolster the sparsity of data
at the lower level of disaggregation with data from other sources or
from other areas or domains are essential to getting estimates with
relatively small standard errors.

The big question is with respect to what distribution are the standard
errors computed. There are three different answers depending on one's
perspective. Sampling statisticians most often wish to take
expectations with respect to the random structure in the sampling
design. At the other extreme are Bayesians for whom the variability is
an inherent part of the stochastic model structure for the phenomenon
of interest, for example, unemployment or crime. And in the middle are
model-based likelihood statisticians. My argument is that in the
context of small area estimation the design-based statisticians were
singularly unsuccessful until they emulated Bayesian ideas of smoothing
and borrowing strength, but even then they have insisted on averaging
with respect to the sampling design, with arguments about robustness of results.

Jiang and Lahiri (\citeyear{Jiang2006}) suggested that the problem goes back almost a
millennium to the eleventh century, but interest in formal statistical
estimation for small areas is a relatively recent phenomenon and much
of the recent literature can be traced to a seminal article by Fay and
Herriot (\citeyear{Fay1979}) who used the James--Stein ``shrinkage'' estimation ideas
to carry out small area estimation in a frequentist manner. Given the
close relationship between such techniques and empirical Bayesian
estimation (e.g.,
see  Efron and Morris, \citeyear{Efron1973}) and mixed linear models,
it is a relatively small leap to the use of fully Bayesian methodology.
But the evolution toward such methodology documented by Jiang and
Lahiri has been relatively slow and marked by a general resistance in
statistical agencies to use models to begin with, let alone Bayesian
formulations; for example, see the descriptions of small area
estimation methodology in the book by Rao (\citeyear{Rao2003}), and contrast it with
the Bayesian hierarchical formulations in the work of Ballin, Scanu and Vicard
(\citeyear{Ballin2005}) and Trevisani and Torelli (\citeyear{Trevisani2004}).

\textit{Census adjustment:} What is remarkable about the ascendency
of the small area estimation methodology in the United States is that
many of those who argued for its use opposed the use of essentially the
same ideas for census adjustment for differential undercount in the
1980s and 1990s. The basic component of census adjustment in these
debates was the use of the now standard capture-recapture methodology
for population estimation (e.g.,
see  Bishop, Fienberg and Holland,
\citeyear{Bishop1975}, Chapter~6), methodology that has its roots in Laplace's method of
ratio estimation. Because a second count (the recapture) in a census
context cannot reasonably be done for the nation as a whole, methods
that utilize a sample of individuals were introduced in 1950 and to get
small area estimates of population, that is, for every block in the
nation, Ericksen and Kadane (\citeyear{Ericksen1985}) proposed the use of a Bayesian
regression model for smoothing. Being fully Bayesian was especially
important because of the sparseness of the data at their disposal for
adjustment, based on a sample from the Current Population Survey. As we
noted above, Freedman and Navidi (\citeyear{Freedman1986}) opposed the use of this
methodology as did Fay and Herriot's colleagues at the US Census
Bureau, at least in part on its use of models with unverifiable
assumptions, and precisely because the shrinkage approach embedded in
the methodology borrowed strength across state boundaries to get
sufficiently tight estimates of error.

Ericksen, Kadane  and Tukey (\citeyear{Ericksen1989}) presented\break a~more refined version of
the technical arguments looking back to the 1980 census, as well as
ahead to the 1990 census. For the 1990 census, the US Census Bureau
essentially proposed the use of a frequentist approach that had similar
structure, at least in spirit, to that proposed for 1980, and this was
possible only by increasing the size of the sample used for adjustment
purposes by an order of magnitude. This plan was opposed largely on
political grounds as well as by Freedman and colleagues who continued
to object to the role of statistical models in the estimation
procedure. A similar controversy ensued as planning for the 2000 census
progressed with components for adjustment as well as sampling for
nonresponse followup, and ultimately the Supreme Court stepped in and
interpreted the Census Act as banning the use of sampling for this
purpose. Anderson and Fienberg (\citeyear{Anderson1999}) and Anderson et al. (\citeyear{Anderson2000})
provided extensive details on the 1990 and 2000 adjustment
controversies. While American politicians have eschewed the use of
Bayesian and non-Bayesian adjustment techniques, statistical agencies
in several other countries, such as Argentina, Australia  and the
United Kingdom, have implemented similar methodology, although with
little emphasis on its Bayesian motivation.

%s4 ###
\section{Election Night Forecasting}\label{sec4}

In the United States the use of statistical forecasting of election
outcomes based on early reported returns began in the early 1950s. The
CBS television network employed one of the early computers, the \mbox{UNIVAC},
and the statistician Max Woodbury developed a regression-style model
that was used successfully to predict the outcome of the 1952
presidential election. By 1960, computers had become a major tool of
the US television networks in support of their election night
coverage. Everything was based in some form or another on the
150,000$+$
precincts where votes were cast across the US, and attention focused
on subsets of ``key'' precincts, chosen in different ways by the three
major networks, and on early access to precinct results. The following
description draws upon that in the article by Fienberg (\citeyear{Fienberg2007}).

In 1960, the RCA Corporation which owned the NBC television network,
hired CEIR, a statistical consulting firm, to develop a rapid election
night~pro\-jection procedure. CEIR consultants included Max Woodbury, and
a number of others including John Tukey. Computers were still large,
expensive and slow, and much of what Max Woodbury had done for CBS
still had to be done by hand. Data of several types were available:
past history (at various levels, e.g., county), results of polls
preceding the election, political scientists' predictions, partial
county returns flowing in during the evening, and complete results for
selected precincts. The data of the analyses were, in many cases,
swings from sets of base values derived from past results and from
political scientists' opinions. It turned out that the important
problem of projecting turnout was more difficult than that of
projecting candidate percentage. Starting with the 1962 congressional
election, Tukey assembled a statistical team to develop the required
methodology and to analyze the results as they flo\-wed in on election
night. Early members of the
team included Bob Abelson, David Brillinger, Dick Link, John Mauchly
and David Wallace who joined for the \citeyear{Mosteller1964} primaries. From 1962 through
1966, they were consultants to RCA and they interacted with the
political scientists and one-time Census Bureau official Richard
Scammon who had his own methodology using a collection of key precinct
results.

David Brillinger (\citeyear{Brillinger2002}) recalled: ``Tukey sought `improved' estimates.
His terminology was that the problem was
one of `borrowing strength'.'' There is a~remarkably close resemblance
between this metho\-dology and that used for small area estimation. The
novel feature in the election night context comes from the nature of
the sparsity---because estimation was based on early reported returns.
The methodology is now recognizable as hierarchical Bayesian with the
use of empirical Bayesian techniques at the top level. Data flowed in
with observations at the precinct (polling place) level and were
aggregated to county level, and then to the state as a whole.
Subjective judgment was used in the choice of the subsets of ``key''
precincts and prior distributions were typically based on the results
of prior state elections with the choice being made subjectively to
capture the political scientists' best judgment about which past
election most closely resembled the election at hand. As early returns
arrived at the computing central command facility, a team of
statisticians reviewed the actual distribution of early returns across
the state to check for anomalies in light of special circumstances and
political practices.

And estimates that really mattered were those at the state level since
the model was used for statewide elections for governor and senate
positions as well as for presidential elections where state outcomes
play a crucial role. Two models were used: one for projecting turnout
and the other for projecting the actual percentage difference
(``swing'') between Democratic and Republican candidates. The occasional
rise of serious independent candidates led to model extensions and
empirical complications.

Brillinger went on to note: ``Jargon was developed; for example, there
were `barometric' and `swing-o-metric' precinct samples. The procedures
developed can be described as an early example of empirical Bayes. The
uncertainties,
developed on a different basis, were just as important as the point
estimates.'' The variance calculations appeared nowhere in the
statistical literature and thus they had to be derived and verified by
members of the team. This was at about the same time as David Wallace
was working with Frederick Mosteller on their landmark Bayesian study
of \textit{The Federalist Papers}, which was published in 1964.
Tukey's attitude to release of the techniques developed is worth
commenting on. Brillinger recounted how,
on various occasions, members of his ``team'' were asked to give talks
and write
papers describing the work. When Tukey's permission was sought, his
remark was
invariably that it was ``too soon'' and that the techniques were
``proprietary'' to
RCA and NBC. With Tukey's death in 2002, we may well have lost the
opportunity to learn all of the technical details of the work done 40~years earlier.

Tukey's students and his collaborators began to use related ideas on
``borrowing strength,'' for example, in the National Halothane Study of
anesthetics (Bunker  et al.,  \citeyear{Bunker1969}) and for the analysis of contingency
table data (e.g.,
see  Bishop, Fienberg  and Holland, \citeyear{Bishop1975}). All of
this before the methodology was described in somewhat different form by
I.~J.~Good in his 1965 book and christened as ``hierarchical Bayes'' in
the classic \citeyear{Lindley1972} paper by Dennis Lindley and Adrian Smith. The specific
version of hierarchical Bayes in the election night model remained
unpublished, although in an ironic twist, something close to it
appeared in a paper written by one of David Wallace's former students,
Alastair Scott, and a colleague, Fred Smith (\citeyear{Scott1969,Scott1971}), who were
unaware of any of the details of Wallace's work for NBC and who
developed their approach for different purposes! Several other
hierarchical Bayesian election night forecasting models have now been
used in other countries, for example, see the work of Brown, Firth and
 Payne
(\citeyear{Brown1997}) and Bernardo and Gir\'on (\citeyear{Bernardo1992}).

The methods described here were in use at NBC through the 1980
presidential elections. Other networks used different methodology and
the statisticians who worked for the Tukey team were quite proud of
their record of early and more accurate calls of winners than those
made by the other networks, especially in close elections. With
Reagan's landslide presidential victory in 1980, the results were
seemingly better captured by exit polls and from 1982 onward NBC
switched to the use of exit polls in competition and then in
collaboration with the other television networks. See the article by
Fienberg (\citeyear{Fienberg2007}) for further details and a number of the recent
controversies regarding exit poll forecasting and reporting.

%s5 ###
\section{Bayesian Methodology and the US Food and Drug Administration}\label{sec5}

Traditional randomized clinical trials, evaluated with frequentist
methodology, have long been viewed as the bedrock of the drug and
device approval system at the US Food and Drug Administration (FDA).
Over the~past couple of decades the drug companies and some members of
the US Congress have been critical of the lengthy FDA review
processes that have resulted and the enormous expense associated with
bringing drugs and medical devices to market. The statistical
literature has also produced Bayesian randomized design alternatives
(e.g.,
see  Spiegelhalter, Freedman   and Parmar \citeyear{Spiegelhalter1994}; Berry, \citeyear{Berry1991,Berry1993,Berry1997}; Berry
and Stangl,  \citeyear{Berry1996}; Simon, \citeyear{Simon1999}), as well as ethical critiques of
traditional frequentist trials (e.g., see Kadane, \citeyear{Kadane1996}). Aside from
the actual interpretation of the outcomes in a Bayesian framework,
these and other authors have argued that the Bayesian approach can
provide faster and more useful clinical trial information in a wide
variety of circumstances in comparison with frequentist methodology.

Bayesian designs and analyses are part of an increasing number of
premarket submissions to FDA's Center for Devices and Radiological
Health (CDRH). This initiative, which began in the late 1990s, takes
advantage of good prior information on safety and effectiveness that is
often available for studies of~the same or similar recent generation
devices. In 2006, CDRH issued draft guidelines for the use of Bayesian
statistics in clinical trials for medical
devices (FDA, \citeyear{FDA2006}) and these were finalized in 2010 (FDA, \citeyear{FDA2010}).
Previous regulatory guidelines have mentioned Baye\-sian methods briefly,
but this was the first broadly circulated specific document focusing on
Bayesian methodologies. The guidelines do, however, place considerable
onus on the drug companies who wish to present Bayesian studies,
largely because of justifiable concerns over selective use of data from
within studies and the reporting of results.

As the guidelines make clear, Bayesian formulations and methods can
improve the assessment of new drugs and devices by incorporating expert
opinion, results of prior investigations, both experiments and
observational studies, and synthesizing results across concurrent
studies. There are sections that emphasize the importance of
hierarchical models and the different roles for exchangeability, for
example, among patients within trials and among trials. We quote from
the final guidelines on the role of prior information:

\begin{quote}

We recommend you identify as many  sour\-ces of good prior information as
possible. The evaluation of ``goodness'' of the prior information is
subjective. Because your\break trial will be conducted with the goal of FDA
approval of a medical device, you should present and discuss your
choice of prior information with FDA reviewers (cli\-nical, engineering
and statistical) before your study begins.

Possible sources of prior information include:
\begin{itemize}
\item clinical trials conducted overseas,
\item patient registries,
\item clinical data on very similar products,
\item pilot studies.
\end{itemize}

\end{quote}

The guidelines go on:

\begin{quote}

Prior distributions based directly on data from other studies are the
easiest to evaluate. While we recognize that two studies are never
exactly alike, we nonetheless recommend the studies used to construct
the prior be similar to the current study in the following aspects:
\begin{itemize}
\item protocol (endpoints, target population, etc.), and
\item time frame of the data collection (e.g., to ensure that the
practice of medicine and the study populations are comparable).
\end{itemize}
In some circumstances, it may be helpful if the studies are also
similar in investigators and sites.
Include studies that are favorable and nonfavorable. Including only
favorable studies creates bias. Bias, based on study selection may be
evaluated by:
\begin{itemize}
\item the representativeness of the studies that are included, and
\item the reasons for including or excluding each study.
\end{itemize}
Prior distributions based on expert opinion rather than data can be
problematic. Approval of a device could be delayed or jeopardized if
FDA advisory panel members or other clinical evaluators do not agree
with the opinions used to generate the prior  (pages~22--23).

\end{quote}

The FDA guidelines include examples of Bayesian studies that have met
agency review standards. Two examples are:
\setcounter{footnote}{1}\begin{example}[(T-Scan)]
\hspace*{-12pt}\footnote{
\href{http://www.accessdata.fda.gov/scripts/cdrh/cfdocs/cfTopic/pma/pma.cfm?num=p970033}%
{http://www.accessdata.fda.gov/scripts/cdrh/cfdocs/cfTopic/}
\href{http://www.accessdata.fda.gov/scripts/cdrh/cfdocs/cfTopic/pma/pma.cfm?num=p970033}%
{pma/pma.cfm?num=p970033}.}\hspace*{10pt}
T-scan 2000 is a device to be used as an adjunct to mammography for
patients with equivocal results. The FDA was presented with an ``intended-use'' study of 74 consecutive biopsies in Italy.
The company combined the results with those from a prospective double
blind study at seven centers compared T-scan to T-scan plus mammography
for 504~patients, and the results from a ``targeted'' study of 657 biopsy
cases at two centers in Israel using a Bayesian multinomial logistic
model. It was able to demonstrate effectiveness in intended use context
where there was insufficient information to demonstrate effectiveness.
The prior was chosen to smooth the zero counts but to be relatively
diffuse. The device was approved for this use as a consequence in 1999.
\end{example}
\begin{example}[(Inter Fix)]
\hspace*{-12pt}\footnote{
\href{http://www.accessdata.fda.gov/scripts/cdrh/cfdocs/cfTopic/pma/pma.cfm?num=p970015}%
{http://www.accessdata.fda.gov/scripts/cdrh/cfdocs/cfTopic/}
\href{http://www.accessdata.fda.gov/scripts/cdrh/cfdocs/cfTopic/pma/pma.cfm?num=p970015}%
{pma/pma.cfm?num=p970015}.}\label{exam2}\hspace*{10pt}
Inter Fix is an implant device for spinal fusion procedure for
patients with degenerative disc disease and back pain. There were data
available for 139 patients in randomized clinical trial, with 77
treated and 62 controls. There were also 104 nonrandomized subjects treated.
An interim analysis was performed based on a Bayesian predictive model
for the future success rate of the device, although most of the other
analyses reported appear to be frequentist in nature. The device was
approved in 1999 as well.
\end{example}

CDRH statisticians have been exploring and lecturing on important
lessons learned in the course of the Bayesian initiative for the
design, conduct  and analysis of medical devices studies such as the
two outlined here.

Although the two studies described above made use of the pooling of
evidence, in many ways the key benefit of Bayesian methods is the
ability it offers to change the study's course when the welfare of
subjects is at stake---using what is known as adaptive randomization.
As Don Berry has argued:

\begin{quote}

In a multiyear frequentist study, new patients will have
the same chance of being enrolled in either group, regardless of
whether the new or old drug is performing better. This approach can put
patients at a disadvantage. A Bayesian model, on the other hand, can
periodically show researchers that one arm is outperforming the other
and then put more new volunteers into the better arm. (Don Berry quo\-ted
in Beckman,  \citeyear{Beckman2006})

\end{quote}

\noindent
As is the case in other applications, at the FDA the main criticism of
the Bayesian approach is the difficulty associated with the choice of
the prior. Spiegelhalter, Freedman  and Parmar (\citeyear{Spiegelhalter1994}) stressed the use of different
forms of priors such as reference priors, ``clinical'' priors,
``skeptical'' priors, and enthusiastic priors. The FDA guidelines
clearly argue against ``subjective'' expert opinion, but as we know from
other settings the likelihood function is often at least as subjective
as is the prior and hierarchical Bayesian structures impose substantial
constraints on the prior and thus the posterior even when one uses
``diffuse'' distributions on the parameters at the highest levels of the
hierarchy! Moreover, when one is drawing upon previous studies, there
is always an issue of how much ``weight'' these should receive in the
prior, especially if the previous studies did not involve randomization
as in Example~\ref{exam2}.

Unfortunately, as these ideas move to other parts of the FDA they are
not without controversy. While we were completing this article, a new
controversy over a specific drug made news. Vasogen Inc. announced that
on Friday, March 14, 2008 it had an initial teleconference with the FDA
to discuss and clarify the recent FDA comments regarding the use of a
Bayesian approach for ACCLAIM II, a clinical trial which is being
planned to support an application for US market approval of the
Celacade\texttrademark\ System for the treatment of patients with New
York Heart Association Class II heart failure.\footnote{FDA deals
blow to Vasogen's heart treatment, Reuters, March~3, 2008.} Oversight of
the drug approval had shifted from CDRH---which had issued the
guidelines for use of Bayesian me\-thods---to the FDA Center for
Biologics Evaluation and Research (CBER), which has adopted a far more
cautious approach. How such issues will work themselves out remains to
be seen.

Another place at the FDA where Bayesian methodology has recently come
into vogue is in the post-approval surveillance of drugs and devices,
especially with regard to side effects. DuMouchel (\citeyear{DuMouchel1999}) discussed
hierarchical Bayesian models for analyzing a~very large frequency table
that cross-classifies adverse events by type of drug used.
Madigan et al. (\citeyear{Madigan2010}) described a more elaborate, large-scale approach
to the analysis of adverse event data gathered via spontaneous
reporting systems linked to claims databases.

$\!\!$It is worth noting that Bayesian methods have~been used in innovative
ways to study the combination~of evidence across studies on matters
directly before~the FDA. On the advice of an expert panel, the FDA in
2004 put a ``black-box'' warning---its highest \mbox{warning} level---on
antidepressants for pedia\-tric use especially among teenagers. The
panel's~ad\-vice was based not on actual suicides, but on indi\-cations
that suicidal thoughts and behaviors increa\-sed in some children and
teens taking newer selective serotonin reuptake inhibitor (SSRI)-type
anti\-depressants. Kaizar et al. (\citeyear{Kaizar2006}) later addressed the combination
of evidence using a hierarchical Bayesian meta-analytical approach.
They concluded that the evidence supporting a causal link between
SSRI-type antidepressant
use and suicidality in children is weak. This will clearly be evidence
that the FDA will need to consider when it next reviews this issue, as
it surely will, because of subsequent observational studies that
suggest teen suicides have increased considerably despite a substantial
decrease in the use of antidepressants (e.g., see Gibbons et al., \citeyear{Gibbons2007}).

Finally we note the extensive applications of a~ran\-ge of Bayesian
methods in the related matters of health technology assessment as
described by Spie\-gelhalter et al. (\citeyear{Spiegelhalter2000}) and Spiegelhalter (\citeyear{Spiegelhalter2004}).

%s6 ###
\section{Confidentiality and the Risk--Utility Trade-off}\label{sec6}

Protecting the confidentiality of data provided by individuals and
establishments has been and continues to be a major preoccupation of
statistical agencies around the world. Over the past 30 years,
statisticians within and outside a number of major agencies have worked
to cast the confidentiality problem as a statistical one, and over the
past decade this effort has taken on substantial Bayesian overtones as
the focus has shifted to the trade-off between risk associated with
protection of confidentiality and the utility of databases for
different kinds of statistical analyses. See the articles in the book
by Doyle,  Theeuwes and Zayatz (\citeyear{Doyle2001}) for a broad review of the literature as it stood
about a decade ago.

Some of the earlier confidentiality literature focused on the
protection of data against intruders or ``data snoopers'' and Fienberg,
Makov  and Sanil (\citeyear{Fienberg1997}) proposed modeling intruder behavior (and thus
protection against it) using a subjective Baye\-sian ``matching'' model;
cf. the discussion of Bayesian ``matching'' methods in the book by
D'Orazio, Di Zio and Scanu (\citeyear{D2006}). In \citeyear{Duncan2001}, Duncan et al. suggested a Bayesian
approach to the risk--utility trade-off problem, which was later
generalized in the context of a formal statistical decision theory
model by Trottini and Fienberg (\citeyear{Trottini2002}) and implemented in illustrative
form by Dobra, Fienberg  and Trottini (\citeyear{Dobra2003}) in the context of
protecting categorical databases.

More recently, Ting, Fienberg  and Trottini (\citeyear{Ting2008}) contrasted their
method of random orthogonal matrix masking with other microdata
perturbation me\-thods, such as additive noise, from the Bayesian
perspective of the trade-off between disclosure risk and data utility.
This work has yet to be adopted by statistical agencies, but related
Bayesian modeling in the same spirit by Franconi and Stander (\citeyear{Franconi2002}),
Polettini and Stander (\citeyear{Polettini2004}), Rinott and Shlomo (\citeyear{Rinott2007})  and Forster and
Webb (\citeyear{Forster2007}) has been done in close collaboration with those in agencies
in Israel, Italy  and the United Kingdom.

One other Bayesian approach to confidentiality protection which has
already seen successful penetration into US statistical agencies is
based on the method of multiple imputation approach due originally to
Donald Rubin and proposed by him for application in the context of
protecting confidentiality in \citeyear{Rubin1993}. See the article by Fienberg, Makov
and Steele (\citeyear{Fienberg1998}) for a related proposal. The basic idea is simple
although the details of the implementation can be complex. We want to
replace the actual confidential data by simulated data drawn from the
posterior distribution of a model that captures the relationships among
the variables to be released. Since these ``sampled units'' are
synthetic and do not actually correspond to original sample members,
proponents claim that the resulting data protect confidentiality by
definition---others point out that synthetic people may be close enough
to ``real'' sample members for there still to be problems of possible
re-identification. The method of multiple imputation allows one to
generate\vadjust{\eject} multiple synthetic (imputed) samples from the posterior and to
use these samples to produce estimates of variability that have a~%
frequentist interpretation. Raghunathan, Reiter  and Rubin (\citeyear{Raghunathan2003}) and
authors of a number of subsequent articles described the formalisms of
the methodology as well as extensions involving only partially imputed
data. Because statistical agencies in the US were already
experimenting with multiple imputation to deal with missing value
problems, a number of them have recently experimented with this
technology for confidentiality protection as well. Since the
methodology works for fairly general classes of prior distributions it
could utilize, at least in principle, prior information from multiple
sources as well as expert judgment.

%s7 ###
\section{Climate Change and Its Abatement}\label{sec7}

By now there is hardly a literate person who has not heard about global
warming and the dire consequences predicted if we do not change our
behavior regarding the emission of greenhouse gases and aerosols. The
following statements are typical and come from a report to the US
Senate by Thomas Karl (\citeyear{Karl2001}), a senior official in the National Oceanic
and Atmospheric Administration:
\begin{itemize}
\item The natural ``greenhouse'' effect is real, and is an essential
component of the planet's climate process.
\item Some greenhouse gases are increasing in the atmosphere because of
human activities and increasingly trapping more heat.
\item The increase in heat-trapping greenhouse gases due to human
activities are projected to be amplified by feedback effects, such as
changes in water vapor, snow cover, and sea ice.
\item Particles (or aerosols) in the atmosphere resulting from human
activities can also affect climate.
\item There is a growing set of observations that yields a collective
picture of a warming world over the past century.
\item It is likely that the frequency of heavy and extreme
precipitation events has increased as global temperatures have risen.
\item Scenarios of future human activities indicate conti\-nued changes
in atmospheric composition throughout the 21st century.
\end{itemize}
These and similar conclusions have been shared with the public by the
Intergovernmental Panel on Climate Change (IPCC) and the US National
Academy of Sciences--National Research Council through a\vadjust{\eject} series of
committee reports. Many of the statements are backed up by elaborate
statistical assessments and modeling and over the past decade this work
has taken on an increasingly Bayesian flavor. There have also been
challenges to many of these statements, despite what the ``global
warming'' proponents describe as increasingly strong empirical support.
See, for example, the report by Wegman, Scott  and Said (\citeyear{Wegman2006}) for a
statistical critique of some recent modeling efforts.

%
%f1 ###
\begin{figure*}

\includegraphics{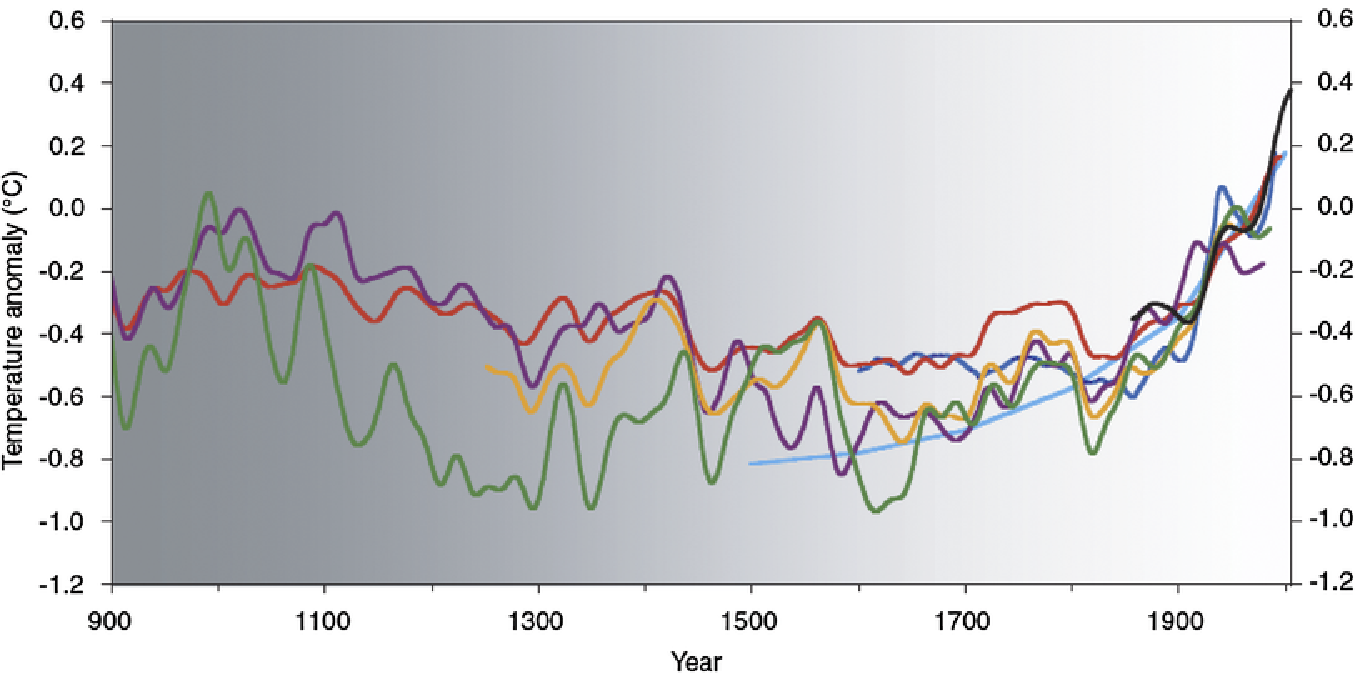}

\begin{tabular}{@{}ll@{}}

\includegraphics{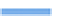}

Borehole temperatures (Huang et al., \citeyear{Huang2000})&
\includegraphics{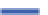}
 Glacier lengths (Oerlemans, \citeyear{Oerlemans2005b})\\

\includegraphics{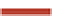}
 Multiproxy (Mann and Jones, \citeyear{Mann2003a})& \includegraphics{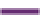} Multiproxy (Moberg et al., \citeyear{Moberg2005a}) \\

\includegraphics{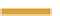}
 Multiproxy (Hegerl et al., \citeyear{Hegerl2006})& \includegraphics{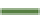} Tree rings (Esper et al., \citeyear{Esper2002a})  \\

\includegraphics{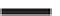}
 Instrumental record (Jones et al., \citeyear{Jones2001})
\end{tabular}
\caption{Smoothed reconstructions of large-scale (Northern Hemisphere
mean or global
mean) surface temperature variations from six different research teams
are shown along with
the instrumental record of global mean surface temperature. Source:
Figure S-1, National Research Council (\protect\citeyear{National2006}), page~2. Reproduced with
permission.}\label{fig:NRCClimateChange}
\end{figure*}

In Figure~\ref{fig:NRCClimateChange} we reproduce an example of the
temperature reconstruction for the past 2000 years based on multiple
sources prepared by a panel from the National Research Council (\citeyear{National2006});
see also National Academy of Sciences (\citeyear{National2008}). One thing that is obvious
from this figure is the convergence of the data sources for the past
150 years, from the start of the industrial revolution, showing
temperatures increasing substantially throughout recent times---this is
global warming! What is also clear is the uncertainty associated with
these reconstructions going back further in time---this is indicated by
the shading in the background of the figure, with darkness associated
with greater uncertainty; cf. the article by Chu (\citeyear{Chu2005}).

The precise trajectory of the recent increases in temperature clearly
has substantial uncertainty\break across the data sources and models and it
would surprise few of us to learn that projections from these data can
vary dramatically. This has recently been the focus of intensive
Bayesian analysis by a number of authors around the world; see, for
example, the articles by Min and Hense (\citeyear{Min2006,Min2007}), and especially
work in the United States by Berliner, Levine and Shea (\citeyear{Berliner2000}), Tebaldi et al.
(\citeyear{Tebaldi2005})  and Sanso, Forest  and Zantedeschi (\citeyear{Sanso2008}).

Tebaldi, Smith  and Sans\'o (\citeyear{Tebaldi2010}) described a way to combine an
ensemble of computer simulation mo\-del results and projections and
actual observations via hierarchical modeling in order to derive
posterior probabilities of temperature and precipitation change at
regional scale. They considered the ensemble of computer models as
being drawn from a~superpopulation of such models, and used
hierarchical Bayesian models to combine results and compute the
posterior predictive distribution for a new climate model's projections
along with the uncertainty to be associated with them. For a related
discussion about assessing the uncertainties of projections, see the
article by Chandler, Rougier and Collins (\citeyear{Chandler2010}).

Whether in the context of this work, or in many other efforts to
forecast future temperatures, Baye\-sian and non-Bayesian, almost all
modeling efforts agree that temperatures will continue to rise. Where
the principal disagreements come in is ``by how much'' and ``what would
be the impact by various strategies for abatement.''

It is worth noting that subjective Bayesian methods were proposed for
use in climate modeling as early as \citeyear{Hobbs1997} by Hobbs and the prominence of
Baye\-sian arguments is due not only to statisticians working in this
area but also to climate modeling specialists such as Schneider (\citeyear{Schneider2002}),
who has noted:

\begin{quote}

For three decades, I have been debating alternative solutions for
sustainable development with thousands of fellow scientists and policy
analysts---exchanges carried out in myriad articles and formal
meetings. Despite all that, I readily confess a lingering frustration:
uncertainties so infuse the issue of climate change that it is still
impossible to rule out either mild or catastrophic outcomes, let alone
provide confident probabilities for all the claims and counterclaims
made about environmental problems.

Even the most credible international assessment body, the
Intergovernmental Pa\-nel on Climate Change (IPCC), has refu\-sed to
attempt subjective probabilistic~esti\-mates of future temperatures. This
has~for\-ced politicians to make their own guesses about the likelihood
of various degrees of global warming. Will temperatures in 2100
increase by 1.4 degrees Celsius or by 5.8? The difference means
relatively adaptable changes or very damaging ones\ldots

So what then is ``the real state of the\break world''? Clearly, it isn't
knowable in traditional statistical terms, even though subjective
estimates can be responsibly offered. The ranges presented by the IPCC
in its peer-reviewed reports give the best snapshot of the real state
of climate chan\-ge: we could be lucky and see a mild effect or unlucky
and get the catastrophic outcomes.

\end{quote}

\noindent
The IPCC assessment builds on formal and informal use of subjective
assessments of the evidence. There is in fact now a tradition in this
field of expert elicitation of expert judgments; for example, see the
articles by Morgan and Keith (\citeyear{Morgan1995}), Keith (\citeyear{Keith1996})  and Zickfeld
et al. (\citeyear{Zickfeld2007}).

%s8 ###
\section{Disability Among the Elderly}\label{sec8}

In the United States, there are no official government surveys of
disability and how it is changing over time, but the National Institute
on Aging (NIA) has funded, with support of other government agencies,
two major longitudinal surveys that capture information on disability
and link it to other data---the Health and Retirement Survey (HRS) and
the National Long Term Care Survey (NLTCS). The original cohort for the
NLTCS was surveyed in 1982 and there have been subsequent~wa\-ves in
1984, 1989, 1994, 1999 and 2004. The NLTCS has been managed by a
university-based organization since the late 1980s, but actual data
collection has been carried out by the US Census Bureau. Considerable
interest in the NLTCS has focused on a series of measures of disability
know as ``Activities of Daily Living'' (ADLs) and ``Instrumental
Activities of Daily Living'' (IADLs), especially for those in the sample
exhibiting some dimension of disability on a~screener question.
Erosheva, Fienberg  and Joutard (\citeyear{Erosheva2007}) studied a cross-sectional
version of 16 binary ADLs and IADLs, represented in the form of a
$2^{16}$ contingency table using a Bayesian latent variable model that
was developed to be an analogue to the frequentist Grade of Membership
(GoM) model of Manton, Woodbury  and Tolley (\citeyear{Manton1994}), the likelihood
function for which is notoriously problematic.

The Bayesian version of the GoM model utilizes hierarchical modeling
ideas through a layered latent variable structure. Let $x=(x_1, x_2,
\ldots , x_J)$ be a vector of binary manifest variables. The GoM model
is structured around $K$ mixture components (extreme profiles), and it
assigns to each individual a~latent partial membership vector of $K$
nonnegative random variables, $g=(g_1, g_2, \ldots , g_K)$, whose
components sum to~1. By assigning a distribution $D(g)$ to the vector
$g$ and integrating, we obtain the marginal distribution for individual
response patterns in the form of individual-level mixtures. Erosheva,
Fienberg and Joutard explained how to fit this Bayesian GoM model
using MCMC techniques and apply it to the data in the $2^{16}$
contingency table displaying outcomes on the 16 ADLs and IADLs,
treating these different measures of disability as exchangeable, and
thus as if they were independent and drawn from another common
distribution. Airoldi et al. (\citeyear{Airoldi2007,Airoldi2010}) explored related aspects of
model specification and model choice. As with a number of the earlier
examples, the hierarchical latent structure embedded in this modeling
approach is a mechanism for gaining control over what might otherwise
be an unmanageable number of parameters and essential to the success of
the related methods.

This work on disability opens the door to a number of challenging
problems for the Bayesian modeling community. For example:
\begin{itemize}
\item How should a Bayesian working with hierarchical models such as
the Bayesian GoM model incorporate the survey weights that arise from
the sampling scheme of the survey and adjustments for nonresponse?
There is now an extensive literature that provides conflicting advice
on the use of survey weights in the Bayesian framework, but the
hierarchical model complexities bring these issues into somewhat
sharper focus in this setting; for example, see the contrasting
arguments of Fienberg (\citeyear{Fienberg2009}) and Little (\citeyear{Little2009}).
\item Manrique-Vallier and Fienberg (\citeyear{ManriqueFienberg2010}) extended these ideas to
longitudinal latent profiles applied to the six ADLs measured across
all six waves of the survey, and Manrique-Vallier (\citeyear{Manrique2010}) added in
survival and generational effects to address the question of whether
disability is increasing or decreasing over time. He appeared to be
able to capture characteristics that others have addressed using
comparisons across cross-sections for each wave of the survey (see,
e.g., Manton and Gu, \citeyear{Manton2001};   Manton, Gu  and Lamb,  \citeyear{Manton2006}). Scaling
these methods up to the full array of ADLs and IADLs with key
covariates remains a major challenge. This is a matter of considerable
interest to policy planners who are interested in forecasting future
demands on the health-care infrastructure as a result of changes in
long-term disability over time.
\end{itemize}

The Bayesian GoM model is a special case of\break a~much larger class of
mixed membership models that can be used to analyze a diverse array of
data types ranging from text in documents to images, to linkages in
networks, and longitudinal versions may pro\-ve applicable in other
settings beyond the study of disability.

%s9 ###
\section{Conclusion}\label{sec9}

For much of the twentieth century, approaches to the design and
analysis of statistical studies in government settings and public
policy were almost\vadjust{\eject} exclusively descriptive or dominated by the
frequentist approach that followed from the work of Fisher and from
Neyman and Pearson. With the neo-Bayesian revival of the 1950s,
Bayesian methods and techniques slowly began to appear in the public
arena, and their use has accelerated dramatically during the past two
decades, especially with the rise of MCMC methods that have allowed for
the sampling from posterior distributions in settings involving very
large datasets.

In this article, we have attempted to give some examples, both old and
new, of Bayesian methods in statistical practice in government and
public policy settings and to suggest why in most of the cases there
was ultimately little or no resistance to the Bayesian approach. Our
examples have included cen\-sus-taking and small area estimation, US
election night forecasting, studies reported to the US Food and Drug
Administration, assessing global climate change and measuring declines
in disability among the elderly. Their diversity suggests that there is
growing recognition of the value of Bayesian results, and a realization
that the approach deals directly with questions of substantive
interest.

Where there has been controversy, it has largely~fo\-cused on the role of
the choice of prior distributions and the appropriateness of
``borrowing strength''\break across geographic boundaries. Arguments in favor
of the use of ``objective'' priors have done little to stem the
frequentist criticism of Bayesian methods, and typically ignore the
highly subjective aspects of elements on hierarchical structures and
likelihood functions. Through the examples discussed here, we have
tried to convey the fact that a pragmatic Bayesian approach inevitably
includes many subjective elements, although prior distributions may
well draw on data from related settings and have an empirical flavor to
them. Nonetheless, the principal challenge to Bayesian methods that
remains is the need to constantly rebut the notion that frequentist
methods are ``objective'' and thus more appropriate for use in the
public domain.

In other areas of statistical application Bayesian methodology has also
seen a major resurgence and this is especially true in connection with
machine learning approaches to very large datasets, where the use of
hierarchically structured latent variable models is essential to
generating high-quality estimates and predictions.

\section*{Acknowledgments}
Supported in part by NIH Grant R01 AG023141-01 to the
Department of Statistics and Army contract DAAD19-02-1-3-0389 to CyLab
at Carnegie Mellon University and by NSF Grants EIA9876619
and IIS0131884 to the National Institute of Statistical Sciences.

\end{document}